\documentclass[11pt,letterpaper]{article}
\usepackage[letterpaper,margin=1in]{geometry}
\pdfoutput=1
\usepackage{indentfirst}
\usepackage[utf8]{inputenc}
\usepackage{cite,hyperref}
\usepackage[compact]{titlesec}
\usepackage{paralist}
\usepackage{xcolor,xspace,graphicx, listings}
\usepackage{algorithm, algorithmicx, algpseudocode}
\algnewcommand{\LeftComment}[1]{\Statex \(\triangleright\) #1}
\algnewcommand\algorithmicforeach{\textbf{for}}
\algdef{S}[FOR]{ForEach}[1]{\algorithmicforeach\ #1\ \algorithmicdo}
\makeatletter
\renewcommand{\ALG@beginalgorithmic}{\scriptsize}
\makeatother
\algrenewcommand{\alglinenumber}[1]{\scriptsize#1:}
\usepackage[small]{caption}
\author{}
\usepackage{outlines}
\newcommand{\allnotes}[1]{}
\renewcommand{\allnotes}[1]{#1} 
\newcommand{\eat}[1]{}

\DeclareCaptionFormat{myformat}{#1#2#3\hrulefill}
\newcommand{\eg}{{\it e.g.,}\xspace}
\newcommand{\ie}{{\it i.e.,}\xspace}
\title{Certifying Safety when Implementing Consensus}
\date{}
\author{\rm{Aurojit Panda}\\ New York University}
\begin{document}
\maketitle
\noindent\textbf{Abstract:} Ensuring the correctness of distributed system \emph{implementations} remains a challenging and largely unaddressed problem. In this paper we present
a protocol that can be used to certify the safety of consensus implementations. Our proposed protocol is efficient both in terms of
the number of additional messages sent and their size, and is designed to operate correctly in the presence of $n-1$ nodes failing in an $n$ node
distributed system (assuming fail-stop failures). We also comment on how our construction might be generalized to certify other protocols and invariants.
\section{Introduction}
\label{sec:introduction}
Correctly implementing distributed systems remains challenging. As distributed systems have become a crucial part of the infrastructure, underlying many popular services, there has been tremendous interest and work in improving their reliability. Interactive theorem provers such as TLA+~\cite{Yu1999ModelCT}, Coq~\cite{Coq:manual}, Isabelle/HOL~\cite{Nipkow2002APA}, etc., and automated theorem provers such as Dafny~\cite{Leino2010DafnyAA}, IVY~\cite{Padon2016IvySV}, etc. are increasingly used to check the correctness of distributed protocols~\cite{Newcombe2015HowAW} before deployment. However, while these tools can ensure that the protocol meets safety and liveness conditions, they cannot provide similar guarantees for implementations.

In response to this gap between protocol and implementation correctness, several recent projects, \eg IronFleet~\cite{Hawblitzel2015IronFleetPP}, Ivy~\cite{Padon2017PaxosME}, Diesel~\cite{Sergey2017ProgrammingAP}, etc.,   have proposed \emph{extracting} correct implementations from verified protocol specifications. These implementations are correct as long as these tools correctly account for low-level system behavior (\eg the behavior of Unix sockets), and can be incorrect otherwise~\cite{Fonseca2017AnES}. While these approaches are promising they have several shortcomings: first, incorporating low level optimizations, \eg ones designed to take advantage of network connectivity~\cite{Behrens2016DerechoG, Ports2015DesigningDS} or I/O~\cite{Gafni2003DiskP}, is challenging and can require changes to the underlying extraction tool; second, any changes to the protocol, no matter how minor, requires extracting a new implementation since current tools do not support incremental extraction increasing time and effort required for validation and deployment; third, most deployed systems provide functionality beyond the basic protocol, \eg consensus services such as Chubby~\cite{Burrows2006TheC}, ZooKeeper~\cite{Hunt2010ZooKeeperWC}, etc. offer not just a basic consensus protocol but also a key-value interface, support for leases~\cite{Gray1989LeasesAE}, etc. and incorporating these additional mechanisms into protocol specifications is challenging and might even render these protocols inexpressible in some of these tools. As a result of these shortcomings, at present hand written code generally provides more features than automatically generated, correct by construction code, and thus cannot easily be replaced. As a result adopting these techniques remains challenging in practice, and points to a need for additional techniques to ensure correctness in distributed systems.

In this paper we take a different tack and propose a complementary approach which can be used to \emph{certify} (\ie check) safety properties of a running distributed system implementation. Our approach, which is implemented by distributed processes we refer to as \emph{certification agents}, checks whether inputs provided by the implementation satisfy a programmer provided predicate, which encodes the safety condition being checked. In \S\ref{sec:certification} we show that such a framework suffices to check agreement and validity for distributed consensus implementations.

To enable practical deployments, in our work we aim for certification protocols that are \emph{sound} (\ie they correctly identify all cases where a predicate does not hold), \emph{complete} (\ie they do not generate false positives) and \emph{efficient}. In terms of efficiency this work focuses on certification protocols where during a round of certification (\ie one round of checking that the predicate holds) each node sends no more than a constant number of messages, and the size of each message is at most $O(\log n)$ bits (in an $n$ node distributed system). In practice, beyond these asymptotic bounds we aim for protocols which require each node sending no more than a few bytes for each certification round, thus limiting their overheads. We also aim for protocols which exhibit message locality, \ie where nodes only communicate with a few \emph{neighboring} nodes. The later is useful in practice when certifying systems deployed on networks with heterogeneous link speeds. We discuss some additional concern around deployment and use of such a framework in \S\ref{sec:discussion}.

The approach we describe in this paper builds on recent results on how to distribute verifiers in interactive proof systems~\cite{Kol2018InteractiveDP, Naor2018ThePO, Korman2010ProofLS}. In the context of that work (and borrowing terminology from~\cite{Kol2018InteractiveDP}), we aim to find 1-round distributed Merlin-Arthur (1-dMA) protocols and then execute them in a non-interactive setting. In our setting the prover represent the implementation (which provides us with inputs for the predicate) and the predicate itself; while certification agents represent verifiers. There are well known techniques for translating interactive proofs into non-interactive proofs including the well known Fiat-Shamir heuristic~\cite{Fiat1986HowTP}, however a naive application of these techniques to our setting does not result in efficient algorithms, and we therefore use a different construction here. Our problem statement is also similar to the previous work on proof labeling schemes~\cite{Korman2010ProofLS}, however while that work focused on checking properties of the network topology, here we apply this approach to checking safety properties of distributed protocols. We provide a more detailed discussion of related work in \S\ref{sec:related}.

Finally, while this paper focuses on presenting certification protocols for agreement and validity in consensus algorithms, we hope in the future to generalize this construction to support certification of safety properties for other systems. To do so we need to ask what are the limits on the kinds of predicates that can be checked using a distributed protocol. Recent results by Naor et al~\cite{Naor2018ThePO} present a construction for converting any centralized verifier which can be executed in $O(n)$ time on a RAM machine into a distributed verifier, however this construction requires at three rounds of interaction between the prover and the verifier. Converting this to a non-interactive setting at least requires a random oracle (that can provide the same random number to all nodes) and potentially several rounds of messages, which might violate our requirements. At the same time, in developing our protocol for consensus algorithms we found that most boolean predicates and arithmetic predicates can be efficiently translated to such a framework, indicating that such a framework might suffice for safety properties in several other distributed systems. We leave the question of generality to future work, but provide a brief discussion on this question based on our experience thus far in \S\ref{sec:generalization}.

\section{System Model and Requirements}
\label{sec:model-req}
\begin{figure}[t]
    \centering
    \includegraphics[width=0.9\textwidth]{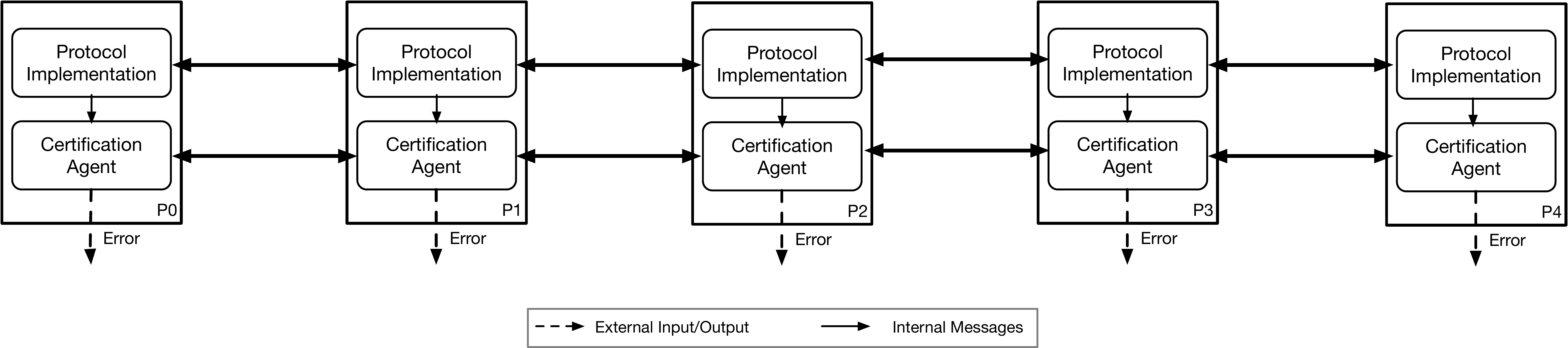}
    \caption{System model: we show a distributed system with $5$ nodes. Each node runs both a process implementing the distributed protocol and a certification agent, which is the focus of this work.}
    \label{fig:design}
\end{figure}
We begin by providing a system model, within which we describe certification protocols and then state requirements for these protocols. We consider distributed systems consisting of $n$ nodes (\eg a 5 node system is shown in Figure~\ref{fig:design}). We assume that a single node can execute several processes which share fate, \ie the failure of one process leads to the failure of all other processes on the same node. We also assume that nodes are connected by a partially synchronous network~\cite{Dwork1988ConsensusIT}, \ie we assume there are bounds on how long any computation can take and for messages to be transferred between nodes. We assume that certification agents can communicate with each other by sending and receiving messages.

In our discussion we assume that each node runs two processes: an \emph{implementation process} running the protocol being certified, and a \emph{certification agent}; thus allowing us to assume fate-sharing between the implementation and the certification agent. Each implementation process can communicate with its \emph{local} certification agent, \ie the certification agent running on the same node. We do not constrain the communication pattern used by implementation processes, nor do we assume visibility into any messages sent or received by these processes. As a result our setting is similar to that used in proof labeling schemes~\cite{Korman2010ProofLS}, with the certification agent in each node being provided an input from the local implementation (referred to as a label in the proof labeling work) and the certification protocol checking whether these inputs satisfy some user-supplied predicate. We discuss our relation to this existing work in greater detail later in \S\ref{sec:related}.

The certificate agents as a whole implement a distributed protocol to check whether the inputs at all nodes satisfy some predicate. We assume that a safety condition is violated when the predicate does not hold, and thus require that at least one (out of $n$) certification agent  signal an \emph{error} whenever the predicate is violated. Note, that the requirement that \emph{at least one} node signal an error is very weak, since we can neither assume that a particular node will always signal errors during predicate violations, nor that all nodes will signal violations. This means that in practice using a certification protocol to check correctness requires a process or administrator to monitor all nodes for error messages, however this allows the use of protocols that can operate in significantly weaker failure models than consensus algorithm (surviving $n-1$ node failures).

A certification round is invoked by the implementation process on a node when it sends a message to the local certification agent. In what follows, for ease of exposition, we assume that implementations at all nodes invoke the certification agent within a short time bound, however we note that this assumption can be relaxed by allowing certificate agents to query the local implementation process for input. We make our description more concrete in the next section.

We require correct certification protocols to provide both \emph{soundness} and \emph{completeness}. We define a certification protocol which evaluates a predicate $P$ to be \textbf{sound} if \emph{at least one node} signals an error on any certification round where predicate $P$ does not hold. We similarly define the certification protocol to be \textbf{complete} if \emph{no node} signals an error in any round where predicate $P$ holds. In this paper we will focus our attention on predicates for which sound and complete certification protocols exist. Our current results do not address the question of what class of predicates can be correctly checked using distributed certification protocols, and we hope to address this question in future work. We discuss this point in greater detail in \S\ref{sec:discussion}.

Beyond correctness we also impose a few efficiency requirements on the certification protocols considered in this paper. These efficiency requirements, which minimize the number and size of messages used for certification, are designed to allow practical deployments of these protocols without requiring operators and developers to add significant network resources to existing systems. Specifically, we require that during each certification round, each node send no more than $O(1)$ messages, and each message sent by the certification protocol be no more than $O(\log n)$ bytes long.  In practice, during failure-free operations, our certification protocols for agreement and validity require sending no more than one message per-node per-certification round, and the messages sent are of constant size. In addition to ensuring that nodes send a fixed number of small messages, we also aim to ensure that our certification protocols require each node to communicate with only a few \emph{neighboring nodes}. The concept of what nodes are neighboring here can be decided on by operators, thus enabling the use of certification in deployments where some nodes might be in the same datacenter, while others might only be accessible over a wide area network and thus communication between different sets of nodes imposes different costs. A key focus of our future work is to design a framework for automatically generating \emph{efficient and correct} certification protocols for a wide range of predicates, which we discuss in \S\ref{sec:generalization}.

\section{Certification of Consensus Protocols}
\label{sec:certification}
In this section we make concrete the notion of distributed certification protocols by developing protocols that can be used to check the correctness of distributed consensus protocols~\cite{Lampson1996HowTB}. We focus on consensus protocols since they have been widely studied: we have theoretical bounds on what a consensus protocol can achieve~\cite{Fischer1983ImpossibilityOD}, a wide range of consensus protocols have been developed~\cite{Lamport2006FastP, Ongaro2014InSO, Dwork1988ConsensusIT}, targetting a range of different network models (asynchronous, synchronous and partially synchronous) and a range of failure models~\cite{Castro1999PracticalBF}. Furthermore, consensus protocols are widely deployed in practical systems~\cite{Burrows2006TheC, Hunt2010ZooKeeperWC}, forming a critical part of how distributed datastores and systems ensure consistency. Errors in the implementation of these systems can thus affect data consistency and the availability of web services, and thus mechanisms for checking and preserving the correctness of these systems are of practical importance. 

\subsection{Safety Properties Considered}\label{sec:certification:invariants}
We consider consensus protocols where each node receives 0 or more \emph{proposals} and all nodes execute a protocol to \emph{agree} (\ie choose) a single proposal that they then output. We built certification proposals for two safety properties that are satisfied by all consensus protocols~\cite{Lampson1996HowTB}:

\medskip 
\noindent\textbf{Agreement:} which requires that all nodes executing a consensus protocol output the same value.\medskip

\noindent\textbf{Validity:} which request that the agreed upon value was proposed. Equivalently, each node receives a set of proposals, and validity requires that a value output by the consensus protocol be contained union of the sets of proposals across nodes. 

\subsection{Certification Protocol}
Here we provide a brief overview of our protocols for checking agreement and validity, which we present in greater detail below. Both protocols represent circuits which are evaluated over inputs distributed across nodes, the agreement circuit checks for cases where nearby nodes disagree on the chosen value, while the validity circuit checks to see whether or any nearby node has a proposal providing validity of the chosen value. To implement these protocols we therefore need a notion of nearby nodes, and we need to arrange nodes in a topology that allows each node to send a message to one or a few neighbors (so we can meet our efficiency goals) while still ensuring evaluation across all nodes. Our approach to doing so is to arrange the nodes in a spanning tree, and then embedding the circuit on this spanning tree. We allow users of our framework to specify how the spanning tree is create, and they can rely on protocols such as Perlman's spanning tree protocol~\cite{Perlman1985AnAF} or any other protocol including ones that consider the underlying physical network over which communication is performed. Since the spanning tree is computed by a user specified protocol, we also certify that the spanning tree computed is in fact correct, \ie it is structurally a tree, and spans all nodes. 

We begin by presenting certification protocols for the spanning tree, agreement and validity in the failure free case. We present our mechanisms for handling failures in \S\ref{sec:certification:fail}, however we note that this mechanism requires little more than recomputing and re-certifying the spanning tree, for which we again call into the user provided algorithm. We now present these protocols in greater detail.

\subsubsection{Certifying a Spanning Tree}\label{sec:certification:structure}
As specified above, our framework is designed to allow users to use an arbitrary protocol for constructing a spanning tree. We require the output from this spanning tree construction to be supplied to each node as a six tuple comprising of:
\begin{compactitem}
    \item A unique ID for the node. We assume that these ID's are totally ordered -- this allows for integer IDs, IDs based on network address, etc.,
    \item A list of IDs for all active nodes in the system.
    \item The ID of the tree's root.
    \item The ID of the node's parent in the spanning tree, or null in case the node is the root.
    \item A list of IDs of the node's children, or null in case the node is a lead.
    \item Finally, the node's distance from the root.
\end{compactitem}

As mentioned above, we need to certify the correctness of this input, for which we rely on techniques which have already described in prior work including the work of Korman et al~\cite{Korman2010ProofLS} and Naor et al~\cite{Naor2018ThePO}. Certifying the correctness of this input requires checking structural properties of the supplied graph, \ie ensuring that the graph is acyclic, has a single root and spans all nodes; and checking that node IDs are unique.

To check that the graph is acyclic and has a single root we have each node send its parent (if any) the ID of the root node and its distance from the root. The parent then ensures that the root node supplied to it as a part of its input is the same as the root node supplied to its children, and that the distance from any of its children to the root is greater than its distance to the root. Checking that a parents root agrees with its children ensures that the supplied graph cannot be a connected forest, \ie it ensures that the supplied graph either has one root or is disconnected. This is because for the graph to have two roots which are connected, there must be sibling nodes which disagree on the ID of the root, in which case at least one of them must also disagree about the root ID with its parent. Checking distances ensures that the graph is acyclic.

Next, we need to check that the graph spans all nodes, this is done by having each node, starting with the leaves send their parent a message containing the number of nodes in the subgraph rooted at that node. For leaf nodes, this simply requires sending $1$, while any non-leaf nodes add up the counts received from their children, and add one to this sum before sending it to their parent. Once the root has computed this value it checks whether the computed value equals $n$, in which case the supplied graph does span all nodes, and reports an error otherwise.

The procedure thus far is sufficient for proving structural properties of the graph, however it cannot prove that nodes have unique IDs. To prove uniqueness we adopt a multi-set equality protocol described in Naor et al~\cite{Naor2018ThePO}, however in adopting this protocol we needed to make minor modifications to allow its use in a setting where nodes cannot interact with a designated centralized prover. We describe this protocol and our changes next.

\subsubsection{Certifying Unique IDs}\label{sec:certification:unique}
\begin{algorithm}
\caption{Protocol for certifying that node IDs are unique}
\label{alg:unique}
\begin{algorithmic}[1]
    \Procedure{CertifyUniqueID}{ }\Comment{The outer function called to check uniqueness of IDs}
    \For {$i:= 0\ ..\ n$}\Comment{Assuming $n$ nodes in the system}\par
        \State \Call{CheckUniqueID}{i}
    \EndFor
    \EndProcedure
    
    \Procedure{CheckUniqueID}{$v$}\Comment{Run at a node to check if all nodes have unique IDs}\par
    \State $msgs \gets recv()$\Comment{Receive messages from all children.}
    \State \Call{CheckSetEquality}{$v$, $id$, $succ\_id$, $msgs$}\Comment{$id$ is the node's ID, $succ\_id$ is the successor ID.}\label{alg:unique:equality}
    \EndProcedure
    
    \Procedure{CheckSetEquality}{$v$, $a$, $b$, $msgs$}\Comment{Check equality between set $A$ and $B$}
    \LeftComment{$a\in A$ is the value from $A$ at the current node.}
    \LeftComment{$b\in B$ is the value from $B$ at the current node.}
    \State $p0 \gets v - a$\label{alg:unique:e0}\par
    \State $p1 \gets v - b$\par
    \ForEach{$m\in msgs$}\par
        \State $p0\gets p0 \times m[0]$\label{alg:unique:e1}
        \State $p1\gets p1 \times m[1]$
    \EndFor
    \If{IsRoot()}\Comment{Is the current node the root}
        \If{$p1 - p0\neq 0$}\Comment{Check if sets differ}
            \State \Call{SignalError}{ }\Comment{Signal an error}
        \EndIf
    \Else
        \State\Call{Send}{parent, (p0, p1)}\Comment{Send a tuple of products to parent}\label{alg:unique:e2}
    \EndIf
    \EndProcedure
\end{algorithmic}
\end{algorithm}
We show pseudocode for a protocol that can be used to certify that IDs are unique in Algorithm~\ref{alg:unique}. The core of this protocol is the multi-set equality protocol proposed in Naor et al~\cite{Naor2018ThePO}, and the only changes are to allow this protocol to be executed in the absence of a random oracle and without a centralized prover. We briefly explain the original protocol before explaining our technique for eliminating the use of a random oracle and some performance trade-offs this offers.

First, in order to use a set equality protocol we need to show that the checking uniqueness of IDs can be reduced to multi-set equality. In \S\ref{sec:certification:structure} we required that IDs have a total order, and that each node be provided with a list of all IDs. We define the successor of an ID $i$ to be the smallest valid ID $i'$ that is greater than $i$. If there is no valid ID larger than some ID $i$, then we define its successor to be the smallest valid ID in the system. This puts node IDs in a ring. Next we consider two sets: the multi-set $I$ of IDs assigned to each node, and the multi-set $S$ of successors to each node's ID. We note that each node is provided $i \in I$ as input, and can use this and the list of IDs to compute the corresponding element $s\in S$. Finally observe that $I = S$ if and only if each node has a unique ID. As a result, we can see that checking uniqueness requires us to use a distributed multi-set equality protocol to check equality between $I$ and $S$ and running a local check to check that each ID in the list of IDs is unique. In Algorithm~\ref{alg:unique} we assume that all nodes have already checked that the provided list does not contain duplicate IDs.

The multi-set equality protocol itself relies on polynomial evaluation to check equality. First we note that given a set $A = \{a_0, a_1, \ldots a_n\}$, where $a_i$ is an integer drawn from some finite field\footnote{To treat elements in this manner we treat fixed length elements as integers represented by their binary representation, or rely on cryptographic hash functions in cases where elements might be variable length.} one can construct and evaluate the polynomial function $F_A(x) = \prod (x - a_i)$ for any point $x$ using an arithmetic circuit embedded in a spanning tree (which we are given as input, and whose structure we have already checked using the procedure described in the previous section). We show an evaluation mechanism for this polynomial on lines~\ref{alg:unique:e0},~\ref{alg:unique:e1} and~\ref{alg:unique:e2} of Algorithm~\ref{alg:unique}. Next, observe that given two degree $n$ polynomial $F_A$ and $F_B$, the polynomial $F_A - F_B$ must either be $0$ at no more than $n$ points, or $F_A$ and $F_B$ must be equal (and thus the difference must be identically 0). This is because in the case where $F_A\neq F_B$, $F_A - F_B$ is a degree $n$ polynomial, and hence has at most $n$ distinct roots. Given this we can now construct polynomials $F_I$ and $F_S$ from sets $I$ and $S$ as defined in the previous paragraph, and then use this polynomial to check set equality.

In Naor et al's work this check is performed by picking points at random from a field of size $n^3$, and evaluating $F_I - F_S$ on these random points. One can bound the probability of any point picked at random being a root of $F_I - F_S$, and hence can evaluate this polynomial difference on a constant number of points to check multi-set equality. However, in our setting we neither assume a random oracle, nor a designated prover who can broadcast a set of random numbers. We resolve this in Algorithm~\ref{alg:unique} by evaluating the difference on the range $\left[0, n\right]$ which consists of $n+1$ points. While this suffices for correctness, it is inefficient requiring each node to send $O(n\log n)$ bits each time it is executed.

However, we note that in practice node IDs rarely change, \eg they are unlikely to change during normal operation or due to node failures.  As a result, this certification process runs infrequently, likely only in cases where new nodes are added, and if we can assume that IDs remain constant for at least $O(n)$ certification rounds, then we find that our relatively simple protocol requires nodes to send out an amortized $O(1)$ bits each time uniqueness has to be certified, and is thus efficient in practice. However, even when ID uniqueness has to be certified more frequently, we can make use of the root of the spanning tree to further increase efficiency, by having the root compute a set of random numbers and then having them propagate through the tree before executing the certification procedure described above. While this increases efficiency, it comes at the cost of increased protocol complexity.

\subsubsection{Certifying Agreement}\label{sec:certification:agreement}
\begin{algorithm}
\caption{Protocol for certifying agreement between nodes.}
\label{alg:agreement}
\begin{algorithmic}[1]
\Procedure{CertifyAgreement}{$d$, $msgs$}\Comment{$d$ is the output (decision) of the local consensus implementation}
\LeftComment{msgs is the set of messages received from the node's children.}
\ForEach{$m\in msgs$}\label{alg:agreement:beg}
    \If{$m\ne d$}\Comment{Check if a child has decided on a different value}
        \State \Call{SignalError}{ }\Comment{Signal an error}\label{alg:agreement:signal}
    \EndIf
\EndFor\label{alg:agreement:end}
\State \Call{Send}{parent, d} \Comment{Send this node's decision to its parent.}\label{alg:agreement:parent}
\EndProcedure
\end{algorithmic}
\end{algorithm}

Having described how during initialization certification agents need to be organized in a spanning tree, and how to certify correctness for this spanning tree, we now turn to certifying the correctness of the consensus protocol itself. We start by describing our protocol for checking agreement, which we show in Algorithm~\ref{alg:agreement}. For ease of exposition, we show only the core of this and the validity certification algorithm, and we assume that the \texttt{CertifyAgreement} procedure is called similarly to the \texttt{CheckSetEquality} procedure in Listing~\ref{alg:unique} (line~\ref{alg:unique:equality}), \ie a calling procedure first waits to receive messages from all children, collects these messages in a set $msgs$ and then calls \texttt{CertifyAgreement} with the input from the implementation and the set $msgs$.

The certification protocol simply requires that each node check whether the decision reached by it differs from the decision reached by any of its children (lines~\ref{alg:agreement:beg}---\ref{alg:agreement:end}). If so the node signals an error (line~\ref{alg:agreement:signal}), and independent of whether an error is signalled it forwards its local decision to its parent (line~\ref{alg:agreement:parent}). This protocol ensures \textbf{completeness} since any node signally an error must have observed a case where the value it has decided on differs from the value one of its children have decided on, \ie an error is signalled only when the protocol observes disagreement between two nodes. The protocol provides \textbf{soundness} because equality is transitive, and our use of a spanning tree ensures that we perform a transitive equality check between any two nodes in the system.

In terms of efficiency, each node sends a single message at each round of this protocol. For any consensus protocol where the agreed value is of constant size, this message is also of constant size. However, one could use a consensus protocol to decide on values whose size depends on the number of nodes in the system (\eg in case the consensus protocol is used to decide on a schedule of nodes). In this later case we use a cryptographic hash function to map this arbitrary length value to a constant hash string. Since standard assumptions about cryptographic hash functions imply that two distinct values do not hash to the same value with high probability, our protocol ensures soundness and completeness with high probability in this case. Finally, we note that while this protocol has low message complexity, it requires as many rounds of communication as the height of the spanning tress which can mean requiring $O(n)$ rounds in the worse case. However, we note that a good choice of spanning trees (\eg one with height $1$) can be used to reduce the number of rounds required, and flexibility in choosing the tree can be used in practice to optimize communication patterns -- \eg reducing the number of messages sent on links connecting multiple data centers.

\subsubsection{Certifying Validity}
\label{sec:certification:validity}
\begin{algorithm}
\caption{Protocol for certifying validity.}
\label{alg:validity}
\begin{algorithmic}[1]
\Procedure{CertifyValidity}{$(d, P)$, $msgs$}\Comment{$d$ is the output (decision) of the local consensus implementation.}
\LeftComment{$P$ is the set of proposals received by the node.}
\LeftComment{msgs is the set of messages received from the node's children.}
\State $c\gets false$ \label{alg:validity:beg}
\ForEach{$p\in P$} \label{alg:validity:local:beg}
    \State $c\gets c\lor (p = d)$
\EndFor \label{alg:validity:local:end}
\ForEach{$m\in msgs$} \label{alg:validity:global:beg}
    \State $c\gets c\lor m$
\EndFor\label{alg:validity:global:end}
\If{\Call{IsRoot}{ }} \Comment{The complete certificate is only visible to the root.}\label{alg:validity:signal}
    \If{$\neg c$}
        \State \Call{SignalError}{ }
    \EndIf
\Else
    \State \Call{Send}{c} \Comment{Send the certificate for the subtree rooted at this node to its parent.}
\EndIf
\EndProcedure
\end{algorithmic}
\end{algorithm}
We present our protocol for certifying validity in Algorithm~\ref{alg:validity}. Certifying validity requires that we find at least one node where the value decided by the consensus algorithm was proposed (\S\ref{sec:certification:invariants}). Our procedure for certifying validity requires that one also check agreement using the protocol in \S\ref{sec:certification:agreement}, however these protocols can be run concurrently. For generality, we also assume that a single node may receive an arbitrary number of proposals.

Given these assumptions our algorithm starts by first checking if local information is sufficient to certify validity, \ie did the current node receive a proposal with the same value as the one that was decided (lines~\ref{alg:validity:local:beg}---\ref{alg:validity:local:end}). Next, we use logical disjunction (\ie the \texttt{or} function) to combine the boolean value of this local certificate with certificates received from the node's children (lines~\ref{alg:validity:global:beg}---\ref{alg:validity:global:end}). Each node then forwards this combined certificate to its parent. Note that the combined certificate can be used to check whether any node in the subtree rooted at the current node can certify validity. As a result the root's certificate signifies whether any node in the entire network can certify validity, and as a result the root signals an error (Line~\ref{alg:validity:signal} in case its computed certificate is false.

Since no node except for the root signals errors in this protocol, \textbf{completeness} is preserved as long as the root's computed certificate is correct. \textbf{Soundness} is similarly a consequence of the root's computed certificate being correct. We observe that starting from the leaf of the tree a local certificate is true if and only if a node has a proposal that proves validity. Furthermore, for any non-leaf node $n$ the combined certificate is true if and only if node $n$ has seen a protocol proving validity or there is a path from some node $n'$ to $n$, where $n'$ has seen such a proposal. As a result for the computed certificate to be true at the root, there must be a node that is contained in the spanning tree (thus ensuring there is a path from the node to the root) which has a proposal proving validity, and must be false otherwise. This ensures safety and completeness.

The efficiency of this protocol is similar to the agreement protocol in \S\ref{sec:certification:agreement}, except that all messages carry a single boolean value.

\subsection{Handling Failures in Fail Stop Environments}\label{sec:certification:fail}
Thus far in our exposition we have no considered cases where nodes might fail. We designed our current protocol to deal with fail-stop errors. We observe that a failed node can disrupt the structure of the spanning tree our protocols rely on. As a result when a node fails we need to recompute a new spanning tree, a task for which we can use existing protocols including the well-known spanning tree protocol~\cite{Perlman1985AnAF}. Once a new spanning has been computed, we rely on the certification protocol in \S\ref{sec:certification:structure} to check structural properties of the spanning tree. We assume node IDs do not change as a result of failures, and thus do not execute \texttt{CertifyUniqueID} (\S\ref{sec:certification:unique}).

Once a new spanning tree has been computed, we can now again use the protocols described in \S\ref{sec:certification:agreement} and \S\ref{sec:certification:validity} to check agreement and validity. However, in this case failures might result in cases where our check for agreement is not sound (\ie we do not detect a case where agreement is violated) or in cases where our check for validity is not complete (\ie we return an error even when a valid proposal exists). This is because we do not have visibility into the state at failed nodes, so in cases where a node communicates an incorrect decision to an external entity and then fails, our agreement check will not in fact observe a violation and hence will not signal an error. Similar, if the failing node is the only node in the system to have seen a proposal that validates the selected value, then our validation protocol cannot observe this proof and hence will generate an error where non exists. To address this problem, we modify our definitions of soundness and completeness to only consider data that is visible to certification agents during a certification round. Specifically, we now define a certification protocol which evaluates a predicate $P$ to be \textbf{sound} if \emph{at least one node} signals an error on any certification round where predicate $P$ does not hold for inputs from non-faulty nodes. We similarly define the certification protocol to be \textbf{complete} if \emph{no node} signals an error in any round where predicate $P$ holds for inputs from non-faulty nodes.

Finally, we conjecture that this gap in soundness and completeness is fundamental, and \emph{no certification protocol} can ensure soundness and correctness while considering the data at failed nodes, \emph{without} assuming knowledge of the distributed system implementation.

\subsection{Putting it All Together}
\begin{algorithm}
\caption{Overall protocol for certifying consensus.}
\label{alg:overall}
\begin{algorithmic}[1]
\Procedure{CertifyConsensus}{($agreement$, $proposals$)}\label{alg:overall:consensus}
\State $m\gets recv()$\Comment{Receive messages from children}
\State \Call{CertifyAgreement}{$agreement$, $m$}
\State $m\gets recv()$\Comment{Receive messages from children}
\State \Call{CertifyValidity}{($agreement$, $proposals$)}
\EndProcedure

\Procedure{DetectFailure}{ }\label{alg:overall:fail}
\State \Call{RecomputeSpanningTree}{ }
\State \Call{CertifySpanningTreeStructure}{ }\Comment{Check structural properties of the spanning tree.}
\EndProcedure

\Procedure{Initialize}{ }\label{alg:overall:init}
\State \Call{RecomputeSpanningTree}{ }
\State\Call{CertifySpanningTreeStructure}{ }
\State\Call{CertifyUniqueID}{ }
\EndProcedure
\end{algorithmic}
\end{algorithm}

Finally, we summarize the protocol executed by certification agents for consensus in Algorithm~\ref{alg:overall}. Certification agents run the initialization function (line~\ref{alg:overall:init} when the system is started, whenever node IDs change, or whenever a new node is added to the system; they respond to failures by recomputing the spanning tree and checking spanning tree structure (\texttt{DetectFailure}, line~\ref{alg:overall:fail}); and respond to certification requests from the implementation by certifying agreement and validity (\texttt{CertifyConsensus}, line~\ref{alg:overall:consensus}). We note that while for ease of exposition we separate checking for agreement and validity into separate phases in Algorithm~\ref{alg:overall}, they are trivially combined into one phase where each node combines messages from both protocols into a single tuple.

\section{Certificates for Other Protocols}
\label{sec:generalization}
In the previous section we presented a certification protocol for consensus. We now turn our attention to generalizing this certification framework so it can both be used to certify additional distributed protocol. What is desirable in generalizing this framework would be to find a procedure that takes as input programmers supplied predicates, and the generates an efficient certification protocol for checking these predicates. We can then rely on existing work (such as the code generation tools described in \S\ref{sec:introduction}) to generate code for certification agents.

Thus, an important question in generalizing the certification framework in this way is to determine the types of predicates for which we can automatically generate efficient protocols. This is an open question, that we are currently working on, but the construction in \S\ref{sec:certification} provides some early hints on what might be possible. In particular, the protocol for checking validity in \S\ref{sec:certification:validity} indicates that boolean formulas whose inputs are distributed across nodes of distributed system can be efficiently evaluated; the certificate for spanning tree structure in \S\ref{sec:certification:structure} show so can arithmetic circuits including ones which either count the number of elements or add up values at different nodes; and the protocol for checking agreement in \S\ref{sec:certification:agreement} shows that we can also efficiently check equality for a fixed number of values. 

More generally, we also see some hope in the recent results from Naor et al~\cite{Naor2018ThePO}, which describes a \emph{compiler} that can be used to convert any centralized verifier that can run in $O(n)$ time on a RAM machine into a 3-round distributed protocol that can be executed in the interactive proof setting. It is unclear whether this result can be translated to our scenario in general, since the construction itself depends on the prover to execute the verifier algorithm while recording a trace, after which the verifiers use a multi-set equality protocol to check the correctness of this trace. In our case we do not assume a distinguished prover, and assuming otherwise would require the prover to have information about all nodes in the implementation. However, despite the challenges of adapting this approach to our setting, we continue to investigate how to apply these results to our setting. Finally, it is also unclear whether predicates for all safety properties can be expressed as $O(n)$ algorithms, \eg when certifying conflict-serializability~\cite{Papadimitriou1979TheSO} in a distributed database, one may need a predicate over the set of transactions, requiring the use of an algorithm with higher asymptotic complexity.

\section{Discussion}
\label{sec:discussion}
Next, we briefly discuss a few additional questions and concerns about certification, before presenting a survey of related work and concluding.

\noindent\textbf{Using Certification Protocols} Thus far in the paper we have not addressed the question of how implementors and operators might use certification protocols. In particular, since on predicate violation any arbitrary node can output an error message, it seems unclear how one might use this message, \eg using such a mechanism as a means to revert erroneous processing would require any error message to be broadcast to all nodes which might itself require execution of an atomic broadcast protocol. Our original intention with certification protocols was to provide a mechanism that could warn developers and operators of erroneous runs, allowing them to then use offline processes to analyze and fix the error. The protocols described thus far suffice for this purpose. In future work we hope to develop additional mechanisms that might be used by implementations to automatically react to certification errors.

\noindent\textbf{Choice of failure models} Our choice of the fail-stop failure model might appear strange in the context of certification, where we are trying to check whether an implementation is bug free, since it requires that we place some trust in the implementation so that inputs to the certification agent are correct. In general we chose the fail-stop model for simplicity, and also because we assume that implementation are not malicious, merely buggy. In this non-malicious setting we assume that implementations do not deliberately send incorrect inputs to the certification agent. Extending our framework for a more realistic byzantine setting requires us to consider clients, who consume the output from the original distributed system, and is left to future work.

\section{Related Work}
\label{sec:related}
We have already discussed some of the related work in earlier section of the work, particularly to work on distributing verifiers in interactive proof systems~\cite{Naor2018ThePO, Kol2018InteractiveDP} which our protocols build on.

The work that is closest to the problem we address is the work on proof labeling scheme first initiated by Korman et al~\cite{Korman2010ProofLS}. The main goal of this proof labeling work was to certify that network configuration met some boolean predicate, \eg certifying that nodes were arranged in a spanning tree or in a clique, nodes had been assigned colors correctly to prove the graph was 3-colorable, etc. Much of the focus this work has been on determining bounds on the label size (which corresponds to the size of the input from the implementation in our model). While our work builds on some of the construction in the proof labeling literature, we differ in two significant ways: first, the class of predicates we consider are different -- our focus is on allowing distributed systems to repeatedly certify their behavior and not on checking topological properties of the network; second, in our setting where certification is run repeatedly failure handling is an important concern to address. To the best of our knowledge, prior work on proof labeling schemes have not considered failures.

There are also close analogs to our certification method and to a proposal from Varghese and Lynch~\cite{Varghese1991SelfstabilizationBL} for developing self-stabilizing algorithms~\cite{Dijkstra1974SelfstabilizingSI}. In this proposal, nodes perform local checks of correctness and execute a protocol to fix any detected violations. Certification serves a similar purpose to the local checks used in this work, however, while our certification protocol could be used in a similar setting, this is not our primary goal.

There is also a large body of work focused on runtime monitoring and verification of distributed systems (see~\cite{Francalanza2017RuntimeVF} for a survey). Similar to our work, these efforts aim to detect and signal cases where the behavior of a distributed system deviates from some specified behavior. However, different from our work, these efforts do so by collecting consistent logs from across the system and analyzing these logs. Consistent log collection is enabled by including vector clocks or other ordering information in messages, this approach is also adopted by logging systems such as Dapper~\cite{dapper} and XTrace~\cite{Fonseca2007XTraceAP}. Compared to certification protocols, these approaches increase the size of each message, adding significant overheads.

\section{Conclusion}
There is significant interest in ensuring correctness for distributed systems, and as a result there has been significant recent work addressing this problem. Much of this work has focused on mechanisms for building distributed systems that are \emph{correct-by-construction}. While the results in this area have been promising, this approach to correctness requires abandoning existing implementations and implicitly trusting the tools that generate these correct-by-construction implementations. In reality both these requirements are hard to meet -- existing implementations usually provide richer functionality than new correct-by-construction implementations, and have often been extensively tested to meet correctness and performance requirements in deployments. In this work we suggest a different approach, where runtime certification is used to efficiently check the correctness of distributed system implementations. We have shown that such certification protocols can be used to check agreement and validity for consensus protocols, and we are now working on determining what set of safety properties are amenable to being checked using certification protocols.

\section{Acknowledgements}
We thank Mooly Sagiv, Scott Shenker, Eylon Yogev and Moni Naor for several early discussions on this work, and help with formulating these ideas. We also thank Changgeng Zhao and Mike Walfish for comments and suggestions on the current write-up. This work was funded in part by an early career faculty award from VMware Research.

\bibliography{assertion}
\end{document}